\documentstyle[12pt,a4,cite,rotate,epsf]{article}

\newcommand{\nn}{\nonumber}

\newcommand{\Mn}{{\cal M}_n}
\newcommand{\wt}{\widetilde}
\newcommand{\as}{\alpha_{s}}

\newcommand{\IM}{\mbox{\rm Im}}

\newcommand{\GG}{\big<aGG\big>}
\newcommand{\eqn}[1]{(\ref{#1})}

\newcommand{\kev}{\mbox{\rm keV}}
\newcommand{\mev}{\mbox{\rm MeV}}
\newcommand{\gev}{\mbox{\rm GeV}}

\newcommand{\Li}{\mbox{\rm Li}_2}
\newcommand{\rms}{\rm\scriptsize}
\newcommand{\MSb}{{\overline{MS}}}

\newcommand{\smvs}{\vbox{\vskip 8mm}}

\newcommand{\gsim}{~{}_{\textstyle\sim}^{\textstyle >}~}
\newcommand{\lsim}{~{}_{\textstyle\sim}^{\textstyle <}~}
\newcommand{\newsection}[1]{\section{#1}\setcounter{equation}{0}}

\begin{document}


\date{\small (February 1997)}

\author{
{\normalsize\bf Matthias Jamin${}^{1}$ and Antonio Pich${}^{2}$} \\
\ \\
{\small\sl ${}^{1}$ Institut f\"ur Theoretische Physik, Universit\"at
           Heidelberg,} \\
{\small\sl Philosophenweg 16, D-69120 Heidelberg, Germany}\\
{\small\sl ${}^{2}$ Departament de F\'\i sica Te\`orica, IFIC,
 CSIC -- Universitat de Val\`encia,}\\
{\small\sl Dr. Moliner 50, E--46100 Burjassot, Val\`encia, Spain}
}

\title{
{\small\sf
\rightline{IFIC/97--06}
\rightline{FTUV/97--06}
\rightline{HD-THEP-96-55}
\rightline{hep-ph/9702276}
}
\bigskip
\bigskip
{\Huge\bf Bottom quark mass and \begin{boldmath}$\alpha_s$\end{boldmath} \\
from the \begin{boldmath}$\Upsilon$\end{boldmath} system \\}
}

\maketitle
\thispagestyle{empty}

\begin{abstract}
\noindent
The mass of the bottom quark and the strong coupling constant $\alpha_s$ are
determined from QCD moment sum rules for the $\Upsilon$ system. Two analyses
are performed using both the pole mass $M_b$ as well as the mass $m_b$ in
the $\MSb$ scheme. In the pole-mass scheme large perturbative corrections
resulting from coulombic contributions have to be resummed. In the
$\MSb$ scheme this can be avoided by an appropriate choice for the
renormalization scale. For the bottom quark mass we obtain
$M_b=4.60 \pm 0.02\,\gev$ and $m_b(m_b)=4.13 \pm 0.06\,\gev$. Our
combined result from both determinations for the strong coupling is
$\alpha_s(M_Z)=0.119 \pm 0.008$.

\end{abstract}

\newpage
\setcounter{page}{1}


\newsection{Introduction}

The Upsilon system constitutes a rich source of information about the
strong interaction dynamics. The bottom quark mass is sufficiently
heavy for a non-relativistic description to be a good starting point
to analyze the quark-antiquark forces. Thus, potential models ---
including relativistic corrections --- have been successfully used
to understand the spectroscopy of the corresponding mesonic bound states
\cite{bgt:80,bt:81,mar:81}. At the same time, the small size of the hadronic
system makes possible to attempt a short-distance approach. At the relevant
energy scale, $\alpha_s$ is small enough to allow (at least for the lowest
levels) a well-grounded quantum field theory analysis with perturbative
QCD tools, including non-perturbative corrections through the Operator Product
Expansion (OPE) \cite{svz:79}.
While the coulombic part of the $b\bar b$ potential is obviously related
to the static piece of the gluon-exchange interaction, a systematic
short-distance investigation provides a better understanding of the
remaining terms in the heavy quark potential, in terms of fundamental
QCD parameters 
\cite{vol:79,leu:81,vol:82,bb:83,dos:87,kdb:89,kdb:90,ty:94,ty:95,ty:95b,sty:95}.

The short-distance description in terms of quarks and gluons is specially
well suited for inclusive quantities, where no reference to a particular
hadronic bound state is needed. The vacuum polarization $\Pi(q^2)$ induced
by the heavy-quark vector current $\bar b\gamma_\mu b$ is then a key
ingredient to investigate the $J^P=1^-$ $b\bar b$ states. Its imaginary
part can be experimentally determined from the $e^+e^-\to b \bar b$
cross-section:
\begin{equation}
\label{eq:1.1}
R_b(s) \; \equiv \; Q_b^2\,R(s) \; \equiv \;
\frac{\sigma(e^+e^-\to b \bar b)}{\sigma(e^+e^-\to\mu^+\mu^-)}
\; = \; 12\pi Q_b^2\,\IM\,\Pi(s+i\epsilon) \, .
\end{equation}
On the other side, $\Pi(q^2)$ can be calculated theoretically within
the OPE. 
To keep our equations more general, let us consider the vector
current $j_\mu(x)=(\bar Q\gamma_\mu Q)(x)$, $Q(x)$ being a heavy quark
field of mass $M$, specifically the bottom quark in our case:
\begin{equation}
\label{eq:1.2}
\Big(q_\mu q_\nu-g_{\mu\nu}q^2\Big)\,\Pi(q^2) \; = \; i \int \! dx \, e^{iqx}
\,\big< \, T\{\,j_\mu(x)\,j_\nu(0)\}\, \big> \,.
\end{equation}
Throughout this work, $M$ corresponds to the pole of the perturbatively
renormalized propagator, whereas the running quark mass in the $\MSb$
scheme \cite{bbdm:78} renormalized at a scale $\mu$ will be denoted by
$m(\mu)$.

Using a dispersion relation the $n$th derivative of $\Pi(s)$ at $s=0$ can
be expressed in terms of the $n$th integral moment of $R(s)$:
\begin{equation}
\label{eq:1.4}
\Mn \; \equiv \; \frac{12\pi^2}{n!}\left.\Biggl(4M^2\,\frac{d}{ds}
\Biggr)^n \Pi(s)\right\vert_{s=0} \; = \; (4M^2)^n \int\limits_0^\infty \!
ds \, \frac{R(s)}{s^{n+1}} \,.
\end{equation}
For later convenience, the moments $\Mn$ are defined to be dimensionless
quantities.
In addition, it will prove useful to express the moments $\Mn$ in terms of
integrals over the variable $v\equiv\sqrt{1-4M^2/s}$,
\begin{equation}
\label{eq:1.5}
\Mn \; = \; 2 \int\limits_0^1 \! dv \, v(1-v^2)^{n-1} R(v) \,.
\end{equation}
Under the assumption of quark-hadron duality, the moments $\Mn$ can either
be calculated theoretically in renormalization group improved perturbation
theory, including non-perturbative condensate contributions, or can be
obtained from experiment. In this way, hadronic quantities like masses
and decay widths get related to the QCD parameters $\alpha_s$, $m_b$ and
condensates.

For large values of $n$, the moments become dominated by the threshold
region. Therefore, they are very sensitive to the heavy quark mass. This
fact has been exploited since the very first QCD analyses of charmonium
and bottomium \cite{nov:77,nov:78,svz:79,gri:79,bb:81,gmpr:81,rry:81,rry:85}
to extract rather accurate values of $m_c$ and $m_b$. More recently,
it has been suggested by Voloshin \cite{vol:95} that the large-$n$
moments can also be used to get a precise determination of $\alpha_s$
from the existing data on $\Upsilon$ resonances.

The perturbative calculation of the moments contains powers of
$\alpha_s \sqrt{n}$ \cite{nov:77,nov:78,vz:87}, which correspond to the
coulombic contributions; they are associated with the near-threshold
quark-antiquark configurations at typical velocity $v\sim 1/\sqrt{n}$,
so that $\alpha_s \sqrt{n}\sim \alpha_s/v$ is the familiar Coulomb parameter.
At large $n$ these coulombic $(\alpha_s \sqrt{n})^k$ terms should be
explicitly summed up to assure a reasonable convergence of the perturbative
series. By the same token, this large-n behaviour implies a big sensitivity
to the value of $\alpha_s$ \cite{vol:95}.

In ref. \cite{vol:95} the large-$n$ moments $\Mn$ have been studied with
a non-relativistic expansion in powers of $1/n$. Fitting the ${\cal O}(1/n)$
contribution from the sum rules, the analysis of the moments $n=8$, 12, 16
and 20 gave the result: $M_b = 4827\pm 7\,\mev$ and
$\alpha_s^{\overline{\mbox{\rms MS}}}(M_Z) = 0.109\pm 0.001$. The quoted
errors are claimed to include the experimental uncertainties and the
theoretical uncertainty due to subleading $1/n$ terms \cite{vol:95}.

The reasoning of ref. \cite{vol:95} looks indeed very suggestive.
The large-$n$ moments are dominated by the first $\Upsilon$ resonance.
Thus, one is actually starting with a confined bound state. In spite of
that, our ability to make an explicit sum of the coulombic contributions
allows to make an impressive determination of the perturbative coupling.
Obviously, an accurate analysis of the theoretical uncertainties is
called for.

The analysis of ref. \cite{vol:95} was performed at ${\cal O}(\alpha_s)$,
i.e., only the ${\cal O}(1)$ and ${\cal O}(\alpha_s)$ perturbative
contributions to the correlator $\Pi(q^2)$ were included. Therefore,
the scale and scheme dependence of $\alpha_s$ was not under control.
Adopting the $\overline{\mbox{\rm MS}}$ scheme, the running of $\alpha_s$
was included in the Coulomb potential, and used to fix the scale of the
coulombic contributions. For the remaining short-distance perturbative
corrections the BLM prescription \cite{blm:83} was used to justify the
choice $\mu = e^{-11/24} M_b$. Given that, the quoted uncertainty in the
final $\alpha_s$ determination looks rather unrealistic.

In order to make a more reliable analysis one needs to know the size
of the higher-order perturbative corrections. Fortunately, the
${\cal O}(\alpha_s^2)$ contributions to the correlator $\Pi(q^2)$ have
been studied recently \cite{bb:95,cks:96a,cks:96b,hkt:95,chks:96}.
Although a complete analytical calculation of these corrections is
still not available, the present information is good enough to perform
an accurate analysis of the moments $\Mn$.

In this paper, we present a detailed study of the relevant moments, using
all the information on $\Pi(q^2)$ that we are aware of. From the present
experimental data we determine the numerical values of the bottom quark
mass and the strong coupling. Moreover, we perform a thorough analysis
of the associated uncertainties.

The resulting values of $\alpha_s(M_Z)$ and $M_b$ are found to be less
precise than what was claimed in ref.~\cite{vol:95}. Nevertheless, they
still constitute rather accurate determinations. The value of the strong
coupling constant turns out to be in excellent agreement with the more
precise measurements obtained from the $\tau$ hadronic width
\cite{bnp:92,qcd:94,tau:96,jhw:96} or from $Z\to\mbox{\rm hadrons}$ data
\cite{sch:97}. Previous claims \cite{vol:95,shi:96} that low-energy
determinations of $\alpha_s$ result in lower $\alpha_s(M_Z)$ values
than higher-energy ones are thus unfounded. On the other side, our
analysis of the $\Upsilon$ system provides the most precise determination
of the bottom quark mass today.

The known perturbative contributions to the moments are given in Section~2
and the Coulomb resummation is performed in Section~3. The non-perturbative
corrections are discussed in Section~4. Section~5 contains the phenomenological
parameterization, extracted from the present data. The numerical analysis
is presented in Sections~6 and~7, which use the {\it pole mass} and
{\it modified minimal subtraction} schemes, respectively. A short conclusion
is finally given in Section~8.

\newsection{Perturbation theory}

In perturbation theory the vacuum polarization $\Pi(s)$ can be expanded
in powers of the strong coupling constant,
\begin{equation}
\label{eq:3.1}
\Pi^{pt}(s) \; = \; \Pi^{(0)}(s)+a\,\Pi^{(1)}(s)+a^2\,\Pi^{(2)}(s)+\ldots \,,
\end{equation}
with $a\equiv\as/\pi$. Analogously, expansions for $R^{\,pt}(v)$ and $\Mn^{pt}$
can be written.

For the first two terms, analytic expressions are available
\cite{ks:55,schw:73}. Here, we only give $R^{(0)}(v)$ and
$R^{(1)}(v)$. The corresponding formulae for $\Pi^{(0)}(s)$ and
$\Pi^{(1)}(s)$ can for example be found in refs.~\cite{gen:84, bft:93}.
\begin{eqnarray}
R^{(0)} & = & \frac{3}{2}\,v(3-v^2) \,, \label{eq:3.2} \\
\smvs
R^{(1)} & = & 2(1+v^2)(3-v^2)\biggl[\,4\Li(p)+2\Li(-p)+\ln(p)\Big(\ln(1+p)+
2\ln(1-p)\Big)\biggr] \nn \\
& & -\,4v(3-v^2)\Big(\ln(1+p)+2\ln(1-p)\Big) \nn \\
& & -\,\frac{1}{4}\,(1-v)(33-39v-17v^2+7v^3)\ln(p)+\frac{3}{2}\,v(5-3v^2) \,,
\label{eq:3.3}
\end{eqnarray}
where $p\equiv(1-v)/(1+v)$ and $\Li(z)$ is the dilogarithmic function
\cite{lew:81}. The expression for $R^{(1)}$ implicitly includes a factor
$C_F=4/3$.

Using the integral representation \eqn{eq:1.5} for $\Mn$, one finds
the following expressions for the moments:
\begin{eqnarray}
\Mn^{(0)} & = & 3(n+1)\,{\rm B}(5/2,n) \,, \label{eq:3.4} \\
\smvs
\Mn^{(1)} & = & {\rm B}(5/2,n)\,\Biggl\{\,(2n+3)\,{\cal A}_n +
n\,{\cal A}_{n+1} - 4n + 12 \nn \\
& & \hspace{21mm} +\,\frac{6}{n}+\frac{2}{(n+1)}-\frac{4}{(n+2)}-
\frac{6}{(n+3)} \,\Biggr\} \,, \label{eq:3.5}
\end{eqnarray}
with
\begin{eqnarray}
\label{eq:3.6}
{\cal A}_n & = & \frac{4}{3}\,\Biggl\{\,1-\frac{1}{2n}-\frac{3}{(n+1)}-
\frac{3}{2(n+2)} \nn \\
\smvs
& & +\,\biggl(\frac{1}{n}+\frac{1}{(n+1)}\biggr)\sum\limits_{k=1}^{n+1}\,
\Biggl[\, \frac{(n+2)\,{\rm B}(1/2,k)}{k\,{\rm B}(1/2,n)}-\frac{3}{k}+
\frac{2}{(2k-1)}\,\Biggr]\,\Biggr\} \,,
\end{eqnarray}
and ${\rm B}(x,y)$ being Euler's Beta function. The first order moments
$\Mn^{(1)}$ are in agreement with the result found by Generalis \cite{gen:84}.

The second-order vacuum polarization $\Pi^{(2)}(s)$ is still not fully known
analytically. However, the method of Pad\'e approximants has
been recently exploited to calculate $\Pi^{(2)}$ numerically,
using available results at high
energies ($s\rightarrow-\infty$), analytical results for the first seven
moments ${\cal M}_i^{(2)}$ for $i=1,\ldots,7$ and the known threshold
behaviour $R^{(2)}(v)$ for $v\rightarrow0$
\cite{bb:95,cks:96a,cks:96b}. Following the lines of ref.~\cite{cks:96b},
it is convenient to split $\Pi^{(2)}$ according to the colour factors,
\begin{equation}
\label{eq:3.7}
\Pi^{(2)} \; = \; C_F^2\,\Pi_A^{(2)} + C_A C_F\,\Pi_{NA}^{(2)} +
C_F T n_l\,\Pi_l^{(2)} + C_F T\,\Pi_F^{(2)} \,,
\end{equation}
and to treat the four different contributions separately, because they
exhibit different behaviour at threshold. 
$\Pi_A^{(2)}$ and $\Pi_{NA}^{(2)}$ contain the pure
gluonic contributions;
the first term is already present in an abelian theory, whereas
$\Pi_{NA}^{(2)}$ is characteristic of the non-abelian aspects of QCD.
The contributions $\Pi_l^{(2)}$ and $\Pi_F^{(2)}$ arise from diagrams
with internal light and heavy quark loops respectively.\footnote{With
respect to the bottom quark, we consider the up, down, strange and
charm quarks to be massless.} The spectral function $R_l^{(2)}(v)$
is known analytically and $R_F^{(2)}(v)$ receives contributions from
a two-particle cut with threshold at $2M$ which is known analytically
and a four-particle cut with threshold at $4M$ which can be calculated
numerically from a two-dimensional integral \cite{hkt:95,chks:96}.
These results can be used to check the reliability of the Pad\'e
approximation for the moments. We shall not repeat the technicalities
of the calculation of the $\Pi_X^{(2)}$, but refer the reader to
ref.~\cite{cks:96b} for details.

\begin{table}[thb]
\begin{center}
\begin{tabular}{|c||r|r||r|r|r|r||r|}
\hline
n & $\Mn^{(0)}$ & $\Mn^{(1)}$
  & ${\cal M}_{A,n}^{(2)}$ & ${\cal M}_{NA,n}^{(2)}$
  & ${\cal M}_{l,n}^{(2)}$ & ${\cal M}_{F,n}^{(2)}$
  & $\Mn^{(2)}$ \\
\hline
\hline
 1 & 2.4000 & 12.1481 & 11.4197 & 15.9696 & -5.2627 & 1.6358 & 71.2368 \\
 2 & 1.0286 &  7.9822 & 14.3850 & 14.1999 & -4.8914 & 0.6010 & 69.7300 \\
 3 & 0.6095 &  6.0448 & 15.0503 & 12.1448 & -4.2652 & 0.3373 & 64.1861 \\
 4 & 0.4156 &  4.8899 & 15.0403 & 10.5729 & -3.7595 & 0.2238 & 59.1538 \\
 5 & 0.3069 &  4.1158 & 14.7923 &  9.3710 & -3.3623 & 0.1627 & 54.9238 \\
 6 & 0.2387 &  3.5585 & 14.4586 &  8.4283 & -3.0453 & 0.1252 & 51.3801 \\
 7 & 0.1926 &  3.1370 & 14.1001 &  7.6699 & -2.7871 & 0.1003 & 48.3813 \\
 8 & 0.1596 &  2.8067 & 13.7427 &  7.0464 & -2.5727 & 0.0827 & 45.8120 \\
 9 & 0.1351 &  2.5406 & 13.3977 &  6.5244 & -2.3917 & 0.0698 & 43.5843 \\
10 & 0.1163 &  2.3214 & 13.0696 &  6.0804 & -2.2368 & 0.0599 & 41.6318 \\
11 & 0.1015 &  2.1377 & 12.7599 &  5.6980 & -2.1025 & 0.0521 & 39.9043 \\
12 & 0.0896 &  1.9815 & 12.4685 &  5.3648 & -1.9850 & 0.0459 & 38.3628 \\
13 & 0.0799 &  1.8468 & 12.1945 &  5.0718 & -1.8811 & 0.0409 & 36.9772 \\
14 & 0.0718 &  1.7296 & 11.9369 &  4.8118 & -1.7886 & 0.0367 & 35.7233 \\
15 & 0.0650 &  1.6267 & 11.6945 &  4.5796 & -1.7057 & 0.0331 & 34.5821 \\
16 & 0.0592 &  1.5354 & 11.4661 &  4.3706 & -1.6309 & 0.0301 & 33.5379 \\
17 & 0.0542 &  1.4541 & 11.2505 &  4.1816 & -1.5629 & 0.0276 & 32.5780 \\
18 & 0.0499 &  1.3810 & 11.0468 &  4.0097 & -1.5010 & 0.0254 & 31.6919 \\
19 & 0.0461 &  1.3150 & 10.8539 &  3.8527 & -1.4443 & 0.0234 & 30.8707 \\
20 & 0.0428 &  1.2552 & 10.6709 &  3.7086 & -1.3921 & 0.0217 & 30.1070 \\
\hline
\end{tabular}
\end{center}
\caption{\label{tab:1} One-, two- and three-loop perturbative contributions
for the first twenty moments $\Mn$.}
\end{table}
In table~\ref{tab:1}, we give the first twenty moments $\Mn^{(0)}$,
$\Mn^{(1)}$ and $\Mn^{(2)}$, as well as the four contributions to
$\Mn^{(2)}$ separately. The first seven moments for ${\cal M}_{X,n}^{(2)}$
correspond to the analytic expressions of ref.~\cite{cks:96b}, whereas
the moments for $n\geq8$ are our results obtained from the Pad\'e
approximants. In the case of ${\cal M}_{A,n}^{(2)}$ the values arise
from a $[5/4]$ approximant, and the moments ${\cal M}_{NA,n}^{(2)}$,
${\cal M}_{l,n}^{(2)}$ and ${\cal M}_{F,n}^{(2)}$ were calculated from
$[4/4]$ approximants because the constant contribution to $\Pi^{(2)}$
in the limit $v\rightarrow 0$ is unknown. To check the stability of
our results and thus the reliability of the Pad\'e approximation,
either different Pad\'e approximants using the full set of information
can be calculated, e.g. $[4/5]$ or $[6/3]$ in the case of
${\cal M}_{A,n}^{(2)}$, or Pad\'e approximants with one order less
can be constructed by removing one datum. For ${\cal M}_{A,n}^{(2)}$
and ${\cal M}_{NA,n}^{(2)}$ the largest change is found if the seventh
moment is removed as an input datum. In particular, ${\cal M}_{A,20}^{(2)}$
changes by $0.002$ and ${\cal M}_{NA,20}^{(2)}$ by $0.0003$.
The moments ${\cal M}_{l,n}^{(2)}$ and ${\cal M}_{F,n}^{(2)}$ can also
be calculated from the available results for $R_l^{(2)}$ and $R_F^{(2)}$
\cite{hkt:95,chks:96}. In the case of ${\cal M}_{F,n}^{(2)}$, for $n\geq8$,
the moments we are interested in, the contribution of the four-particle
cut can be neglected, and it is sufficient to consider the analytically
available expressions for the two-particle cut. The agreement with the
twentieth moments ${\cal M}_{l,20}^{(2)}$ and ${\cal M}_{F,20}^{(2)}$ as
calculated from the Pad\'e approximants is better than $10^{-6}$ in
both cases. Thus for all moments under consideration the uncertainty
is below 0.02\%, being completely negligible for our application.
We conclude that the method of Pad\'e approximation works sufficiently
well to predict the moments up to at least $n=20$.

Generally, the moments $\Mn^{(2)}$ depend on the renormalization scheme
and scale for the strong coupling constant. The values presented in
table~\ref{tab:1} correspond to $a(M)$ in the $\MSb$ scheme and a
renormalization scale $\mu_a=M$. If the scale $\mu_a$ is varied, the
moments change according to
\begin{equation}
\label{eq:3.8}
\Mn^{(2)}(\mu_a) \; = \; \Mn^{(2)}(M) + \beta_1\Mn^{(1)}
\ln\frac{\mu_a}{M} \,.
\end{equation}
The first coefficients of the $\beta$-function in our conventions
are given in appendix~A.

Besides varying the scale $\mu_a$ at which the coupling constant is
evaluated, in our numerical analysis we shall also use a different definition
of the quark mass. Apart from the pole mass, the sum rules will also be
analyzed in terms of a running $\MSb$ mass $m(\mu_m)$, evaluated at a
scale $\mu_m$. Of course, physical quantities should remain unchanged.
Therefore, the variation in our results originating from changes of
scheme and scale in the coupling and quark mass will give an estimate
of the uncertainty due to higher orders in perturbation theory. It
should already be remarked that we have deliberately chosen different
scales $\mu_a$ and $\mu_m$ in the coupling and mass respectively, in
order to be able to vary them independently.

From the definition~\eqn{eq:1.4},
it is easy to calculate the relations between
the moments defined in terms of the pole mass and those expressed in
terms of a running $\MSb$ mass $m(\mu_m)$:
\begin{eqnarray}
\label{eq:3.9}
\overline{\cal M}_n^{(1)} & = & \Mn^{(1)} + 2n\,r_m^{(1)}\Mn^{(0)} \,, \\
\smvs
\overline{\cal M}_n^{(2)} & = & \Mn^{(2)} + 2n\,r_m^{(1)}\Mn^{(1)} +
n\Big( 2r_m^{(2)}+(2n-1)r_m^{(1)^2} \Big)\Mn^{(0)} \,,
\end{eqnarray}
where $r_m^{(1)}$ and $r_m^{(2)}$ appear in the relation between pole
and running $\MSb$ mass
\begin{equation}
\label{eq:3.10}
m(\mu_m) \; = \; M\,\Big[\,1 + a(\mu_a)\,r_m^{(1)}(\mu_m) +
a(\mu_a)^2\,r_m^{(2)}(\mu_a,\mu_m) + \ldots \,\Big] \,.
\end{equation}
Explicit expressions for $r_m^{(1)}$ and $r_m^{(2)}$ are also given in
appendix~A.

As can be seen from table~\ref{tab:1}, for large $n$ the higher-order
corrections grow with respect to the leading order. At $n=8$ the first
order correction is roughly 120\% of the leading term whereas the
second order contribution is 140\%. At $n=20$ the contributions of
first and second order are 200\% and 340\% respectively. This behaviour
of the perturbation series for large moments is well known
\cite{nov:77,nov:78,vz:87}
and originates from the fact that the relevant parameter in the Coulomb
system is $\as/v$ which leads to a $\as\sqrt{n}$ dependence of the
moments. Thus for higher $n$ the perturbative corrections become
increasingly more important and have to be summed up explicitly
in order for the theoretical expressions to make sense. This Coulomb
resummation will be discussed in the next section.

If, on the other hand, the $\MSb$ mass is used, it is not clear how
a Coulomb resummation could be performed, because now the velocity
$v$ depends on the renormalization scale. However, the radiative
corrections in the $\MSb$ scheme are somewhat smaller than if a pole
mass is used. At $\mu_a=\mu_m=m$ and $n=8$ the first and second order
corrections are $-27$\% and 3\% whereas for $n=20$ they are $-170$\%
and 115\% respectively. This suggests to try to find a scale $\mu_m$
for which the perturbative corrections stay within a reasonable range.
If we require the second-order correction not to exceed 50\%, the
scale $\mu_m$ should lie within $2.7\,\gev\lsim\mu_m\lsim 3.7\,\gev$.
For the numerical analysis it is therefore possible to also exploit
the sum rules in the $\MSb$ scheme if $\mu_m\approx 3.2\pm0.5\,\gev$
is chosen.

\newsection{Coulomb resummation}

Let us first state our Ansatz for the Coloumb resummed spectral function
$R(v)$ and then discuss the different components:
\begin{equation}
\label{eq:4.1}
R(v) \; = \; \Big( 1-4C_Fa+16C_F^2 a^2 \Big)\Big\{\,R^{(0)} + R_C +
\wt R^{(1)} a + \wt R^{(2)} a^2 \,\Big\} \, ,
\end{equation}
with
\begin{eqnarray}
R_C & = & \frac{9}{2}\biggl[\,\frac{x_V}{\big(1-e^{-x_V/v}\big)}-v\,\biggr]
\,, \label{eq:4.2} \\
\smvs
\wt R^{(1)} & = & R^{(1)} + 4C_F R^{(0)} - \frac{9}{4}\,\pi^2 C_F
\,, \label{eq:4.3} \\ \smvs
\wt R^{(2)} & = & R^{(2)} + 4C_F R^{(1)} - \frac{3\pi^4 C_F^2}{8v} -
\frac{9}{4}\,\pi^2 C_F\, r_V^{(1)} \,. \label{eq:4.4}
\end{eqnarray}
Here, $x_V\equiv \pi^2C_F a_V$ and $a_V$ is the effective coupling which
corresponds to the heavy quark-antiquark potential. Expressed in terms
of the $\MSb$ coupling, we have
\begin{equation}
\label{eq:4.5}
a_V(\vec q^{\,2}) \; = \; a(\mu_a)\,\Big[\, 1 + a(\mu_a)\,r_V^{(1)}
(\vec q^{\,2}/\mu_a^2) + a(\mu_a)^2\,r_V^{(2)}(\vec q^{\,2}/\mu_a^2) + \ldots
\,\Big] \,.
\end{equation}
Because $a_V$ is related to the static QCD potential it is independent
of the renormalization scale but it does depend on the three-momentum
transfer between the heavy quark and antiquark. Explicit expressions
for $r_V^{(1)}$ and $r_V^{(2)}$ are given in appendix~B.

The term $R_C$ corresponds to the resummed spectral function resulting
from the imaginary part of the Green function for the QCD Coloumb
potential. It resums the leading $(a/v)^n$ and some of the sub-leading
corrections \cite{vol:79}. The corresponding terms have to be subtracted
from $R^{(1)}$ and $R^{(2)}$. Although the QCD corrections to $R$ are
only known to order $a^2$, we have included the recently calculated
${\cal O}(a^3)$ contribution for $a_V$ \cite{pet:96,pet:97}, in order to
investigate the dependence of our results on higher-order corrections.
This will be discussed further in section~6.
In addition, we have factored out the correction to the
vector current which originates from transversal, hard gluons. To the
known correction ``$-4C_F a$'' we have added a term $16C_F^2 a^2$ in order
not to generate additional corrections of order $a^2$ proportional to
$R^{(0)}$. We shall comment further on this point below. After performing
the Coulomb resummation, the large-moment behaviour of the remaining terms
is much weaker. Let us discuss the different contributions in more detail.

It should be clear from eq.~\eqn{eq:1.5} that the large-$n$ behaviour can
always be inferred from the small-$v$ behaviour of $R(v)$. Expanding
eq.~\eqn{eq:3.4}, we obtain
\begin{equation}
\label{eq:4.6}
\Mn^{(0)} \; = \; \frac{9\sqrt{\pi}}{4n^{3/2}}\,\Biggl\{\,
1-\frac{7}{2^3 n} + \frac{145}{2^7 n^2} - \frac{1645}{2^{10}n^3} +
{\cal O}\biggl(\frac{1}{n^4}\biggr)\,
\Biggr\} \,.
\end{equation}
The small-$v$ expansion of $R^{(1)}$ is given by:
\begin{equation}
\label{eq:4.7}
R^{(1)} \; = \; 3\pi^2 - 24v + 2\pi^2 v^2 + \biggl(16\ln(8v^2)-\frac{148}{3}
\biggr) v^3 - \pi^2 v^4 + {\cal O}(v^5) \,.
\end{equation}
The first two terms are canceled by the additional contributions to
$\wt R^{(1)}$. Therefore, although $\Mn^{(1)}/\Mn^{(0)}$ increases as
$\sqrt{n}$, now
\begin{equation}
\label{eq:4.8}
\frac{\wt \Mn^{(1)}}{\Mn^{(0)}} \; = \; \frac{8\pi^{3/2}}{9\sqrt{n}} -
\frac{16}{3n}\,\biggl[\,\ln\biggl(\frac{n}{2}\biggr)+\gamma_E+\frac{11}{12}
\,\biggr] - \frac{\pi^{3/2}}{n^{3/2}} + {\cal O}\biggl(\frac{1}{n^2}\biggr) \,.
\end{equation}
For the moments $n=8,\ldots,20$ the second term is of the same size as
the first. Thus for the case of interest the large-$n$ expansion is
very badly behaved. In fact, below $n\approx 100$ the ratio
$\wt\Mn^{(1)}/\Mn^{(0)}$ increases and only for $n>100$ the asymptotic
$1/\sqrt{n}$ decrease is approached.

The available analytical results for $R_l^{(2)}$ and $R_F^{(2)}$ allow
to calculate the small-$v$ behaviour for these functions as well:
\begin{eqnarray}
R_l^{(2)} & = & \frac{3\pi^2}{4}\biggl[\ln\frac{4v^2}{(1-v^2)}-\frac{5}{3}
\biggr] + \frac{11}{2}v + \frac{\pi^2}{2}\biggl[\ln\frac{4v^2}{(1-v^2)}-
\frac{17}{3}\biggr]v^2 + {\cal O}(v^3) \,, \label{eq:4.9} \quad \\
\smvs
R_F^{(2)} & = & \Big(22-2\pi^2\Big)v - \biggl(\frac{245}{18}-\frac{4\pi^2}{3}
\biggr)v^3 + {\cal O}(v^4) \,. \label{eq:4.10}
\end{eqnarray}
Again, the first term in eq.~\eqn{eq:4.9} is canceled by the corresponding
piece in the last term of eq.~\eqn{eq:4.4} if we substitute
$\vec q^{\,2}=v^2 s=4v^2M^2/(1-v^2)$ and if the first coefficient of
the $\beta$-function in $r_V^{(1)}$ is evaluated with $n_l$ light
quark flavours. On the other hand, $R_F^{(2)}$ vanishes at threshold
and hence has no contribution which should be resummed in the Coulomb
term. This indicates that consistently the coupling constant in $R_C$
should be evaluated in an effective theory with only $n_l$ active flavours.
To facilitate the numerical analysis, we then prefer to rewrite the
full expression for $\Mn$ in terms of the coupling $a$ defined in the
$n_l$-flavour theory. From the matching relations for $a$
\cite{bw:82,ber:83,lrv:95}, it follows that this just amounts to
using the corresponding $\beta_1$ with $n_l$ flavours in eq.~\eqn{eq:3.8}.

Analogously, all terms of ${\cal O}(1/v)$, ${\cal O}(\ln v^2)$ and
${\cal O}(1)$ for $R_A^{(2)}$ and $R_{NA}^{(2)}$ are canceled in
eq.~\eqn{eq:4.4}, such that $\wt R^{(2)}$ vanishes in the limit
$v\rightarrow 0$. Nevertheless, for these two functions the contributions
of ${\cal O}(v)$ which determine the constant terms in
$\wt{\cal M}_{n,A}^{(2)}/\Mn^{(0)}$ and $\wt{\cal M}_{n,NA}^{(2)}/\Mn^{(0)}$
are not known analytically. Precisely those terms correspond to the
current correction from transversal, hard gluons. In order to obtain
information on the large-$n$ behaviour of the second-order moments, we
can assume an expansion analogous to eq.~\eqn{eq:4.8}, however including
a constant term and fitting this Ansatz to the moments as calculated
from the Pad\'e approximation. We then find
\begin{equation}
\label{eq:4.11}
\frac{\wt{\cal M}_{n}^{(2)}}{\Mn^{(0)}} \; \approx \; 161.1 - \frac{1174.3}
{\sqrt{n}} + 819.9\,\frac{\ln n}{n} + \frac{534.6}{n} + \ldots \,.
\end{equation}
The fit has been determined using moments with $n=20,\ldots,50$, but the
coefficients are rather stable if the number of fit points is changed.
Although the error on the coefficients probably is substantial, it
nevertheless shows that again here the large-$n$ expansion converges
slowly and for the range of $n$ in which we are interested, higher-order
terms have to be included.

From the constant term in eq.~\eqn{eq:4.11}, we can in principle infer
the short distance correction resulting from transversal gluons. However,
in the region of interest, namely for $n=8,\ldots,20$, the ratio
$\wt\Mn^{(2)}/\Mn^{(0)}\approx 40$. This contribution should be added
to the $16C_F^2$ already factorized in eq.~\eqn{eq:4.1}, therefore
further increasing this positive correction. In the work by Voloshin
\cite{vol:95}, the BLM scale setting prescription \cite{blm:83} was
applied to absorb the ${\cal O}(a^2)$ correction in the term
$-4C_F\,a(\mu_a)$ by changing the renormalization scale $\mu_a$, and
it was found that this should be accomplished with the choice
$\mu_a\approx 0.63 M_b$. Because the first and second order terms appear
with different signs, from the explicit calculation we now see that
on the contrary the scale $\mu_a$ should be greater than $M_b$. Because
we keep the ${\cal O}(a^2)$ correction explicitly, there is no need
to evaluate this scale here.

\newsection{Gluon condensate}

Analytical results for the gluon condensate contribution to the massive
vector correlator are available at the next-to-leading order
\cite{svz:79,bro:94}. Adopting the notation of ref.~\cite{bro:94},
the corresponding moments are given by
\begin{equation}
\label{eq:5.1}
{\cal M}_{n,G^2} \; = \; \frac{3\pi^2}{4}\, \frac{\GG}{M^4}\,a_n^V
\Big[\,1+a\, b_n^V \,\Big] \,,
\end{equation}
with
\begin{equation}
\label{eq:5.2}
a_n^V \; = \; -\,\frac{1}{24}\,(n+1)(n+3) B(1/2,n+3) \,,
\end{equation}
and the coefficients $b_n^V$ together with numerical values for
the $a_n^V$ up to $n=21$ are shown in table~\ref{tab:2}.
The coefficients $b_n^V$ depend on the renormalization scheme for the
mass. If the $\MSb$ scheme is used the $b_n^V$ change according to
\begin{equation}
\label{eq:5.3}
\overline b_n^V \; = \; b_n^V + (2n+4)\,r_m^{(1)} \,.
\end{equation}

\vspace{-5mm}
\begin{table}[bht]
\begin{center}
\begin{tabular}{|c|ccccccc|}
\hline
$n$ & 1 & 2 & 3 & 4 & 5 & 6 & 7 \\
\hline
$a_n^V$ & -0.3048 & -0.5079 & -0.7388 & -0.9946 & -1.2730 & -1.5726 & -1.8918\\
$b_n^V$ & 10.4768 & 11.7202 & 12.8494 & 13.8928 & 14.8685 & 15.7888 & 16.6625\\
\hline
\hline
$n$ & 8 & 9 & 10 & 11 & 12 & 13 & 14 \\
\hline
$a_n^V$ & -2.2296 & -2.5851 & -2.9574 & -3.3457 & -3.7495 & -4.1682 & -4.6012\\
$b_n^V$ & 17.4964 & 18.2956 & 19.0643 & 19.8058 & 20.5229 & 21.2179 & 21.8928\\
\hline
\hline
$n$ & 15 & 16 & 17 & 18 & 19 & 20 & 21 \\
\hline
$a_n^V$ & -5.0482 & -5.5087 & -5.9823 & -6.4686 & -6.9674 & -7.4784 & -8.0012\\
$b_n^V$ & 22.5492 & 23.1887 & 23.8124 & 24.4216 & 25.0173 & 25.6002 & 26.1712\\
\hline
\end{tabular}
\end{center}
\caption{\label{tab:2} First- and second-order coefficients for the
gluon condensate contribution to the moments $\Mn$.}
\end{table}

From eq.~\eqn{eq:5.2} it is clear that the relative growth of
${\cal M}_{n,G^2}^{(0)}/\Mn^{(0)}$ is proportional to $n^3$. Therefore,
the non-perturbative contribution grows much faster than the perturbative
moments. In addition, as can be seen from table~\ref{tab:2}, in the
pole-mass scheme at $\mu_a=M$ the next-to-leading order correction
is of the same size or larger as the leading term. Because the perturbative
expansion for the gluon condensate cannot be trusted, we shall restrict
our analysis to a range of $n$ where its contribution to the moments is small
and can be neglected. Using $\GG\approx0.021\,\gev^4$ \cite{bro:94,nar:96},
we find that for $n\leq20$ the contribution from the gluon condensate
to the $b\bar b$ moments
is below 3\%. Thus, we shall restrict our phenomenological analysis to this
range. 

In the $\MSb$ scheme the situation concerning the perturbative
expansion is somewhat better. If we take $\mu_m\approx3.2\,\gev$,
as was discussed at the end of section~2, for $n\leq20$ the next-to-leading
order contribution stays below 70\% of the leading order. This demonstrates
that a determination of the gluon condensate from charmonium should be
performed in the $\MSb$ scheme. Nevertheless, for this work also in the
$\MSb$ scheme we shall keep the restriction to $n\leq20$. A lower limit
on the number of moments will be discussed in the next section.

\newsection{Phenomenological parameterization}

In the preceding sections, theoretical predictions for the spectral function
$R(s)$ and the related moments $\Mn$ have been calculated without further
specifying the actual quark content. For the phenomenological parameterization
of the spectral function we shall now restrict our discussion to the
$b\bar b$ system.

In the narrow-width approximation the contribution to $R_b(s)$ from a
$\Upsilon(kS)$ resonance is given by
\begin{equation}
\label{eq:6.1}
R_{b,kS}(s) \; = \; \frac{9\pi}{\bar\alpha^2}\,\Gamma\big(\Upsilon(kS)\to
e^+e^-\big)\,M_{kS}\,\delta(s-M_{kS}^2) \,,
\end{equation}
where $\bar\alpha$ denotes the running QED coupling evaluated at a scale
around the resonance mass. Because in the Review of Particle Properties
\cite{pdg:96} the electronic widths have been calculated with
$\bar\alpha^2=1.07\,\alpha^2$ where $\alpha=1/137.04$ is the fine
structure constant, we shall use this value accordingly.
In the case at hand the narrow-width approximation is extremely good 
because the full widths of the first three $\Upsilon$ resonances
are roughly a factor $10^{-5}$ smaller than the corresponding masses
and the higher-resonance contributions to the moments are suppressed.
 
Experimentally, the first six resonances have been observed. The measured
masses and electronic widths are collected in table~\ref{tab:3}. For our
numerical analysis the errors on the masses can be safely neglected and
have thus not been listed. Inserting eq.~\eqn{eq:6.1} in the definition
of the moments $\Mn$, eq.~\eqn{eq:1.4}, we obtain
\begin{table}[thb]
\hspace{-12mm}
\vbox{\begin{tabular}{|c|cccccc|}
\hline
 $k$ & 1 & 2 & 3 & 4 & 5 & 6 \\
\hline
$ M_{kS}\,[\gev]$     & 9.460 & 10.023 & 10.355 & 10.580 & 10.865 & 11.019 \\
$\Gamma_{kS}\,[\kev]$ & $1.31\pm0.04$ & $0.52\pm0.03$ & $0.48\pm0.08$ &
$0.25\pm0.03$ & $0.31\pm0.07$ & $0.13\pm0.03$ \\
\hline
\end{tabular}}
\caption{\label{tab:3} Masses and electronic widths of the first six
$\Upsilon(kS)$ resonances.}
\end{table}
\begin{equation}
\label{eq:6.2}
\Mn \; = \; (4M_b^2)^n\,\Biggl\{\,\frac{9\pi}{\bar\alpha^2 Q_b^2}\,
\sum\limits_{k=1}^6 \frac{\Gamma_{kS}}{M_{kS}^{2n+1}} +
\int\limits_{s_0}^\infty \!ds \,\frac{R^{pt}(s)}{s^{n+1}}\,\Biggr\} \,.
\end{equation}
The numerical weight of the heavier resonances in~\eqn{eq:6.2} decreases
strongly for increasing values of $n$. The contribution of the $\Upsilon(5S)$
[$\Upsilon(6S)$] state is 9.5\% [4\%] at $n=0$; 1\% [0.3\%] at $n=10$;
and a tiny 0.08\% [0.02\%] at $n=20$. Therefore, taking $n\gsim 10$,
the uncertainties associated with the contributions of higher-mass states
are very small.

The second term in eq.~\eqn{eq:6.2} accounts for the contributions to
$R_b$ above the sixth resonance and is approximated by the perturbative
continuum. Generally, the continuum threshold $\sqrt{s_0}$ should lie around
the mass of the next resonance, which has been estimated in potential models
\cite{lrwp:90}. For our analysis we shall use $\sqrt{s_0}=11.2\pm0.2\,\gev$.
The lower value for $s_0$ includes the mass of the sixth resonance and should
be a conservative estimate. There is still a contribution missing which stems
from open $B$ production above the $B\bar B$ threshold and below $s_0$.
From the experimental data \cite{cleo:91} we infer that its influence is
small and has been included in the variation of $s_0$.

\newsection{Numerical analysis in the pole-mass scheme}

\begin{figure}[bth]
\vspace{0.1in}
\centerline{
\rotate[r]{
\epsfysize=6in
\epsffile{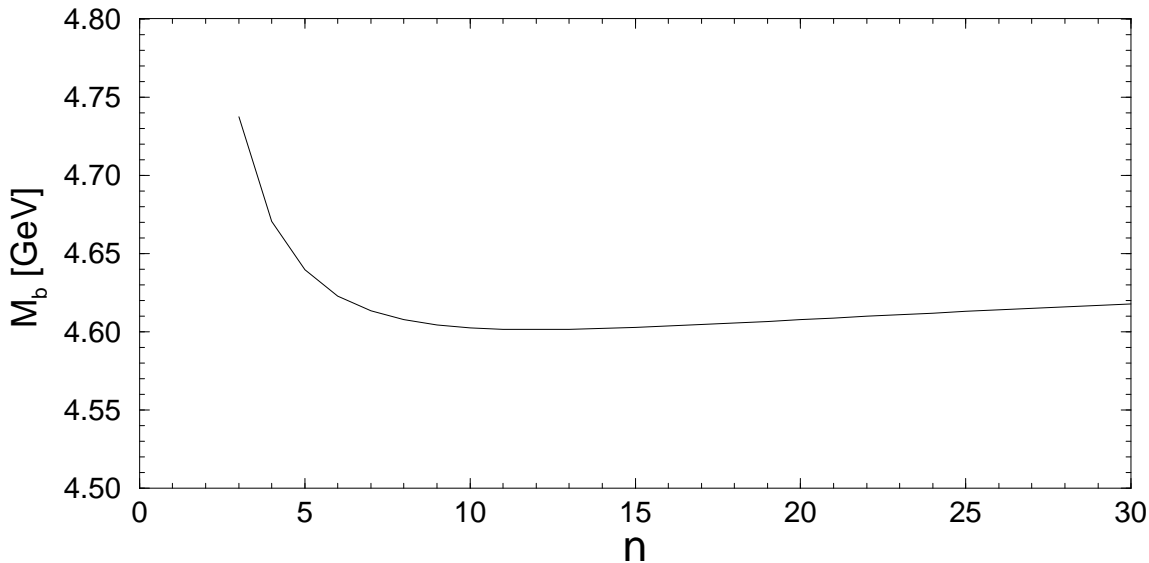}
\vspace{.1in} }}
\caption[]{The $b$ quark pole mass as a function of $n$.\label{fig:1}}
\end{figure}

Quark-hadron duality entails the equality of the theoretical moments
$\Mn^{th}$ presented in sections 2 to 4 and the phenomenological moments
$\Mn^{ph}$ discussed in the previous section. The moments corresponding
to the Coulomb term $R_C$ of eq.~\eqn{eq:4.2} have been calculated from
eq.~\eqn{eq:1.5} by numerical integration. To suppress higher resonances
as well as power corrections, following ref.~\cite{vol:95}, we have
restricted $n$ to the range $n=8,\ldots,20$. Solving the moment sum
rules for $M_b$, we can fit $M_b$ to a constant by varying $M_b$ and
$\alpha_s(M_b)$. The fit has been performed using the program Minuit
\cite{min:94}. For the central set of parameter values our result is
\begin{eqnarray}
M_b & = & 4.604 \pm 0.009 \; \gev \,, \label{eq:6.3} \\
\alpha_s(M_b) & = & 0.2197 \pm 0.0097 \label{eq:6.4} \,.
\end{eqnarray}
The error in these results just corresponds to the statistical error
of the fit. In the fit we have included every second moment to have
less statistical dependence, but the results change very little if
all moments with $n=8,\ldots,20$ or only every fourth moment is used.
In figure~\ref{fig:1} the resulting values for $M_b$ are displayed as
a function of $n$. This illustrates that a constant $M_b$ in the range
$8\leq n\leq 20$ really produces an excellent fit.

\begin{table}[thb]
\begin{center}
\begin{tabular}{|ccc|}
\hline
& $\Delta M_b\;[\mev]$ & $\Delta\alpha_s(M_b)\;[10^{-3}]$ \\
\hline
statistical & $\pm 9.1$ & $\pm 9.7$ \\
${\cal O}(a^3)$ Coulomb & $\pm 7.3$ & $\pm 22.2$ \\
${\cal O}(a^2)$ & $\pm 0.9$ & $\pm 8.2$ \\
scale $\mu_a$ & $\pm^{2.1}_{1.3}$ & $\pm^{28.7}_{19.6}$ \\
continuum & $\pm^{2.6}_{2.0}$ & $\pm^{3.7}_{2.7}$ \\
$\GG$ & $\pm 5.3$ & $\pm 3.6$ \\
$\Gamma_{e^+e^-}$ & $\pm 3.1$ & $\pm 6.7$ \\
\hline
total & $\pm 13.5$ & $\pm 26.9$ \\
\hline
\end{tabular}
\caption{\label{tab:4} Separate contributions to the errors of $M_b$ and
$\alpha_s(M_b)$.}
\end{center}
\end{table}

In the remaining part of this section let us present a detailed discussion
of the errors resulting from the various input quantities. A compilation
of all different contributions to the errors on $M_b$ and $\alpha_s(M_b)$
is summarized in table~\ref{tab:4}. The dominant theoretical uncertainty
is due to the unknown higher-order perturbative corrections. We have
estimated this uncertainty in three different ways. As has been already
remarked in section~3 the relation between the effective coupling $a_V$
in the Coulomb potential and $a^{\MSb}$ is known to ${\cal O}(a^3)$
\cite{pet:96,pet:97}. We can thus include this correction in $R_C$ to see what
the influence on our results is. Although this is not consistent
because the corresponding correction $\wt R^{(3)}$ is not available, it
nevertheless can be taken as an error estimate of higher-order corrections.
A second possibility is a variation of the scale at which $\alpha_s$ is
evaluated. For the result in table~\ref{tab:4} we have chosen the range
$M_b/2 \leq \mu_a \leq 2M_b$. As a final test on the importance of 
higher-order corrections, we can remove the ${\cal O}(a^2)$ term $\wt R^{(2)}$
completely. From table~\ref{tab:4} we observe that including the
${\cal O}(a^3)$ correction to $a_V$ has a much bigger influence than
removing $\wt R^{(2)}$. This is not unexpected because the Coulomb piece
sums up the dominant contributions in the large-$n$ limit. The uncertainty
of the scale dependence is of the same order as the sum of the other two
contributions. For our estimate of the uncertainty resulting from 
higher-order corrections we can now either take the scale dependence or combine
the other two contributions. Adding all three would double count the error,
because the uncertainty in an asymptotic series, such as the perturbative
expansion, is bounded by the size of the last known term. For our final
results, we have chosen to include the errors of varying the Coulomb and
the ${\cal O}(a^2)$ terms.

The error from the continuum contribution has been estimated by varying
$s_0$ in the range $\sqrt{s_0}=11.2\pm0.2\,\gev$. The entry for the gluon
condensate in table~\ref{tab:4} results from removing the gluon condensate
completely and for the uncertainty from the electronic widths we have
varied all widths within the errors given in table~\ref{tab:3}. With
respect to the uncertainty resulting from higher orders all these errors
are small. Adding all errors in quadrature, we arrive at our final
result in the pole-mass scheme:
\begin{eqnarray}
M_b & = & 4.604 \pm 0.014 \; \gev \,, \label{eq:6.5} \\
\alpha_s(M_b) & = & 0.2197 \pm 0.0269 \label{eq:6.6} \,.
\end{eqnarray}
Evolving the strong coupling constant to $M_Z$, we find
\begin{equation}
\label{eq:6.7}
\alpha_s(M_Z) \; = \; 0.1184 \pm^{\;0.0070}_{\;0.0080} \,.
\end{equation}
Our central result is in astonishingly good agreement to the current world
average \cite{sch:97}, although the error turns out to be larger. Further
comments on our results also with respect to the paper by Voloshin
\cite{vol:95} have been relegated to the conclusions.

\newsection{Numerical analysis in the $\MSb$ scheme}

Besides analyzing the moment sum rules exploiting the pole mass $M_b$, in
addition we have investigated the same sum rules in the $\MSb$ scheme. In
contrast to the pole mass, the quark mass in the $\MSb$ scheme depends on
the renormalization scale $\mu_m$. As has been remarked in section~2, to
restrict the ${\cal O}(a^2)$ corrections to a reasonable size, $\mu_m$ should
lie in the range $\mu_m=3.2\pm 0.5\,\gev$. We have refrained from performing
a resummation of the large radiative corrections because now the velocity
$v$ depends on the renormalization scheme and it is not straightforwardly
possible to proceed in analogy to the Coulomb resummation for the pole
mass.

The fitting procedure was performed along the same lines as for the 
pole-mass case.  For the central values of our input parameters, we obtain
\begin{eqnarray}
m_b(m_b) & = & 4.133 \pm 0.002 \; \gev \,, \label{eq:7.1} \\
\alpha_s(m_b) & = & 0.2325 \pm 0.0044 \label{eq:7.2} \,,
\end{eqnarray}
where again the errors are purely statistical. Since it is more standard
to evaluate the running $b$-quark mass at $m_b$, we have evolved our
immediate result $m_b(3.2\,\gev)$ to this scale with the help of the
renormalization group equation.

\begin{table}[thb]
\begin{center}
\begin{tabular}{|ccc|}
\hline
& $\Delta m_b\;[\mev]$ & $\Delta\alpha_s(m_b)\;[10^{-3}]$ \\
\hline
statistical & $\pm 2$ & $\pm 4.4 $ \\
scale $\mu_m$ & $\pm^{33}_{36}$ & $\pm^{36.2}_{\phantom{3}5.8}$ \\
scale $\mu_a$ & $\pm^{49}_{31}$ & $\pm^{25.1}_{28.0}$ \\
continuum & $\pm 1$ & $\pm^{3.5}_{2.6}$ \\
$\GG$ & $\pm 2$ & $\pm 2.3$ \\
$\Gamma_{e^+e^-}$ & $\pm 3$ & $\pm 6.0$ \\
\hline
total & $\pm^{59}_{48}$ & $\pm^{44.9}_{29.7}$ \\
\hline
\end{tabular}
\caption{\label{tab:5} Separate contributions to the errors of $m_b(m_b)$
and $\alpha_s(m_b)$.}
\end{center}
\end{table}
The separate contributions to the theoretical error have been obtained by
performing the same variations as for the pole-mass scheme and have been
listed in table~\ref{tab:5}. The uncertainty from higher-order corrections
is now due to the variation of the scales $\mu_m=3.2\pm 0.5\,\gev$ and
$2.6\,\gev \leq \mu_a \leq 2m_b$. The scale $\mu_a$ should not be taken
lower than roughly $2.6\,\gev$ because otherwise the ${\cal O}(a^2)$
correction $\overline{\cal M}_n^{(2)}$ becomes unacceptably large. Since
in addition to $\mu_a$ for the $\MSb$ scheme we can also vary $\mu_m$,
the resulting uncertainty, especially for $m_b$, is larger than for the
pole mass. Adding all errors in quadrature, we arrive at our final result
in the $\MSb$ scheme:
\begin{eqnarray}
m_b(m_b) & = & 4.13 \pm 0.06 \; \gev \,, \label{eq:7.8} \\
\alpha_s(m_b) & = & 0.2325 \pm^{\;0.0449}_{\;0.0297} \label{eq:7.9} \,.
\end{eqnarray}
Evolving the strong coupling constant to $M_Z$, we find
\begin{equation}
\label{eq:7.10}
\alpha_s(M_Z) \; = \; 0.1196 \pm^{\;0.0102}_{\;0.0080} \,.
\end{equation}

It is gratifying to observe that the resulting values for $\alpha_s(M_Z)$
from the pole-mass and $\MSb$ schemes turn out to be in very good agreement.
This is a further indication that the uncertainty from unknown higher-order
corrections is under control. In addition, our results $m_b(m_b)$ and
$M_b$ for the $b$-quark mass satisfy the relation \eqn{eq:a.4} between the
pole and $\MSb$ mass within the errors. This should be expected because
the relation \eqn{eq:a.4} has been used to rewrite the moment sum rules
in terms of the $\MSb$ mass. Nevertheless, it again shows that variations
due to higher orders are accounted for by our error estimates.

\newsection{Conclusions}

Before we enter a discussion of our findings, let us again summarize the
central results. For the bottom quark mass in the pole-mass as well as
$\MSb$ scheme, we obtain
\begin{eqnarray}
M_b & = & 4.60 \pm 0.02 \; \gev \,, \label{eq:8.1} \\
m_b(m_b) & = & 4.13 \pm 0.06 \; \gev \,, \label{eq:8.2}
\end{eqnarray}
respectively. Combining both determinations of the strong coupling constant
$\alpha_s$, we find
\begin{equation}
\label{eq:8.3}
\alpha_s(M_Z) \; = \; 0.119 \pm 0.008 \,.
\end{equation}
We have not averaged the errors of the two determinations because
they are not  independent.

The bottom quark mass values obtained by us are in good agreement to
previous determinations from QCD sum rules
\cite{rry:81,rry:85,nar:89,dp:92,nar:94}
and a very recent calculation from lattice QCD \cite{gms:96}.
Owing to the big sensitivity of the moment sum rules for the $\Upsilon$
system to the quark mass, and the good control over higher-order $\alpha_s$
corrections, our result is more precise.

Nevertheless, the pole quark-mass value obtained by us is in disagreement
to the result found by Voloshin \cite{vol:95}. In our opinion the
discrepancy is due to the importance of higher ${\cal O}(1/n)$ corrections,
which in ref.~\cite{vol:95}
were either neglected, or numerically fitted
from the sum rules. 
In \cite{vol:95} it was assumed that the
leading order correction goes like $1/n$. 
However, from eqs.~\eqn{eq:4.8} and
\eqn{eq:4.11}, it is clear that they rather behave like $1/\sqrt{n}$.
Besides, we have also demonstrated that for the region of $n$ used in
the analysis, the large-$n$ expansion is not justified.
In addition, the second-order $\alpha_s$ correction was only partially
and partly incorrectly taken into account. Therefore, the scale dependence
of $\alpha_s$ was not under control.

Let us shortly comment on the renormalon ambiguity of the pole mass.
During the last years, it has been realized that beyond perturbation
theory the pole masses for the charm and bottom quarks suffer from
unknown renormalon ambiguities, leading to additional 
theoretical uncertainties
in their determination \cite{bb:94,bsuv:94,ns:94}. On general grounds
this uncertainty has been estimated to be of ${\cal O}(100\,\mev)$. Throughout
our analysis, the pole mass has been defined as the pole of the
perturbatively renormalized quark propagator. 
Our determination \eqn{eq:8.1} might therefore be
subject to additional uncertainties which go beyond perturbation theory
but which we cannot assess in a precise way.

Within our errors the result obtained for $\alpha_s(M_Z)$, eq.~\eqn{eq:8.3},
is compatible with the result by Voloshin~\cite{vol:95}, though, given the
shortcomings of this analysis discussed above, his errors appear to be
largely underestimated. Thus, previous claims of a low value of
$\alpha_s(M_Z)$ from low-energy determinations which could hint to new physics
\cite{vol:95,shi:96} are unfounded. On the other hand, our central value
for $\alpha_s(M_Z)$ is surprisingly close to the current world average
$\alpha_s(M_Z)=0.118\pm0.003$ \cite{sch:97}, although the error is
certainly larger.

The dominant uncertainty for the determination of the $b$-quark mass and
$\alpha_s$ from the $\Upsilon$ system was found to originate from the
dependence on the renormalization scale, or, equivalently, the size of
the as yet unknown higher-order corrections. Improving the error on the
$\alpha_s$ determination will thus only be possible if the full
${\cal O}(\alpha_s^3)$ correction to the moments $\Mn$ is known and if
it turns out to be reasonably small. We hope to return to this question
in the future.

\vspace{6mm} \noindent
{\Large\bf Acknowledgments}
 
\vspace{3mm} \noindent
M. J. would like to thank H. G. Dosch and D. Gromes for helpful discussions.
This work has been supported in part by the German--Spanish Cooperation
agreement HA95-0169.
The work of A.P. has been supported in part by CICYT (Spain) under the
grant AEN-96-1718.

\newpage
\appendix{\LARGE\bf Appendices}

\newsection{Renormalization group functions}

For the definition of the renormalization group functions we
follow the notation of Pascual and Tarrach \cite{pt:84}, except
that we define the $\beta$-function such that $\beta_1$ is positive.
The expansions of $\beta(a)$ and $\gamma(a)$ take the form:
\begin{equation}
\label{eq:a.1}
\beta(a) \; = \; -\,\beta_1 a-\beta_2 a^2-\beta_3 a^3-\ldots \,,
\quad \hbox{and} \quad
\gamma(a) \; = \; \gamma_1 a+\gamma_2 a^2+\gamma_3 a^3+\ldots \,,
\end{equation}
with
\begin{equation}
\label{eq:a.2}
\beta_1 \; = \; \frac{1}{6}\,\Big[\,11C_A-4Tn_f\,\Big] \,, \qquad
\beta_2 \; = \; \frac{1}{12}\,\Big[\,17C_A^2-10C_ATn_f-6C_FTn_f\,\Big] \,,
\end{equation}
and
\begin{equation}
\label{eq:a.3}
\gamma_1 \; = \; \frac{3}{2}\,C_F \,, \qquad
\gamma_2 \; = \; \frac{C_F}{48}\,\Big[\,97C_A+9C_F-20Tn_f\,\Big] \,.
\end{equation}

The relation between pole and running $\MSb$ mass is given by
\begin{equation}
\label{eq:a.4}
m(\mu_m) \; = \; M\,\Big[\,1 + a(\mu_a)\,r_m^{(1)}(\mu_m) +
a(\mu_a)^2\,r_m^{(2)}(\mu_a,\mu_m) + \ldots \,\Big] \,,
\end{equation}
where
\begin{eqnarray}
r_m^{(1)} & = & r_{m,0}^{(1)} - \gamma_1\ln\frac{\mu_m}{m(\mu_m)} \,,
\label{eq:a.5} \\
\smvs
r_m^{(2)} & = & r_{m,0}^{(2)} - \Big[\,\gamma_2+(\gamma_1-\beta_1)\,
r_{m,0}^{(1)}\,\Big]\ln\frac{\mu_m}{m(\mu_m)} + \frac{\gamma_1}{2}\,(\gamma_1-
\beta_1)\ln^2\frac{\mu_m}{m(\mu_m)} \nn \\
\smvs
& & -\,\biggl[\,\gamma_1+\beta_1\ln\frac{\mu_m}{\mu_a}\,\biggr]
r_m^{(1)} \,. \label{eq:a.6}
\end{eqnarray}
The coefficients of the logarithms can be calculated from the renormalization
group and the constant coefficients $r_{m,0}^{(1)}$ and $r_{m,0}^{(2)}$
are found to be \cite{tar:81,gbgs:90,cks:96b}
\begin{eqnarray}
r_{m,0}^{(1)} & = & -\,C_F \,, \label{eq:a.7} \\
\smvs
r_{m,0}^{(2)} & = & C_F^2\biggl(\frac{7}{128}-\frac{15}{8}\zeta(2)-\frac{3}{4}
\zeta(3)+3\zeta(2)\ln 2\biggr)+C_FTn_f\biggl(\frac{71}{96}+\frac{1}{2}\zeta(2)
\biggr) \nn \\
\smvs
& & \hspace{-16mm}+\,C_AC_F\biggl(-\frac{1111}{384}+\frac{1}{2}\zeta(2)+
\frac{3}{8}\zeta(3)-\frac{3}{2}\zeta(2)\ln 2\biggr)+C_FT\biggl(\frac{3}{4}-
\frac{3}{2}\zeta(2)\biggr) \,. \label{eq:a.8}
\end{eqnarray}

\newsection{The effective coupling $a_V$}

In terms of the $\MSb$ coupling the effective coupling $a_V$ is given by
\begin{equation}
\label{eq:b.1}
a_V(\vec q^{\,2}) \; = \; a(\mu_a)\,\Big[\, 1 + a(\mu_a)\,r_V^{(1)}
(\vec q^{\,2}/\mu_a^2) + a(\mu_a)^2\,r_V^{(2)}(\vec q^{\,2}/\mu_a^2) + \ldots
\,\Big] \,,
\end{equation}
where
\begin{eqnarray}
r_V^{(1)} & = & r_{V,0}^{(1)}-\frac{\beta_1}{2}\ln\frac{\vec q^{\,2}}{\mu_a^2}
\,, \label{eq:b.2} \\
\smvs
r_V^{(2)} & = & r_{V,0}^{(2)}-\biggl[\,\frac{\beta_2}{2}+\beta_1 r_{V,0}^{(1)}
\,\biggr]\ln\frac{\vec q^{\,2}}{\mu_a^2} + \frac{\beta_1^2}{4}\ln^2
\frac{\vec q^{\,2}}{\mu_a^2} \,. \label{eq:b.3}
\end{eqnarray}
Like in eq.~\eqn{eq:a.4} the coefficients of the logarithms are determined
by the renormalization group and the constant coefficients $r_{V,0}^{(1)}$
and $r_{V,0}^{(2)}$ are found to be \cite{fis:77,bil:80,pet:96,pet:97}
\begin{eqnarray}
r_{V,0}^{(1)} & = & \frac{31}{36}\,C_A - \frac{5}{9}\,Tn_l \,,\label{eq:b.4} \\
\smvs
r_{V,0}^{(2)} & = & \frac{1}{16}\,\Biggl\{\,C_A^2\biggl(\frac{4343}{162}+
6\pi^2-\frac{\pi^4}{4}+\frac{22}{3}\zeta(3)\biggr)-C_ATn_l\biggl(
\frac{1798}{81}+\frac{56}{3}\zeta(3)\biggr) \nn \\
\smvs
& & -\,C_FTn_l\biggl(\frac{55}{3}-16\zeta(3)\biggr)+T^2n_l^2\,\frac{400}{81}
\,\Biggr\} \,. \label{eq:b.5}
\end{eqnarray}
Here, $n_l=n_f-1$ is the number of light quark flavours.

\newpage

\end{document}